\documentstyle[12pt,psfig]{article}
\begin{document}

\hfill DUKE-CGTP-2000-16

\hfill hep-th/0008191

\vspace*{2in}

\begin{center}

{\large\bf Recent Developments in Discrete Torsion }

\vspace{1in}

Eric Sharpe \\
Department of Physics \\
Box 90305 \\
Duke University \\
Durham, NC  27708 \\
{\tt ersharpe@cgtp.duke.edu} \\

$\,$ \\

\end{center}

In this short article we briefly review some recent developments
in understanding discrete torsion.  Specifically, we give a 
short overview of the highlights of a group of recent papers
which give the basic understanding of discrete torsion.
Briefly, those papers observe that discrete torsion can be
completely understand simply as the choice of action of the
orbifold group on the $B$ field.  We summarize the main points
of that work.

\begin{flushleft}
August 2000 
\end{flushleft}


\newpage

\tableofcontents

\newpage

\section{Introduction}

Historically discrete torsion has been a rather mysterious
aspect of string theory.  Discrete torsion was originally
discovered \cite{vafa1} as an ambiguity in the choice of
phases to assign to twisted sectors of string orbifold partition
functions.  Although other work has been done on the subject
(see, for example, \cite{vafaed,doug1,doug2}), until recently
no work  
has succeeded in giving any sort of 
genuinely deep understanding
of discrete torsion.  In fact, discrete torsion has sometimes been
referred to has an inherently stringy degree of freedom, without
any geometric analogue.

Recently, however, a purely mathematical understanding of discrete
torsion was worked out \cite{dt1,dt2,dt3,cdt,hdt}.
In a nutshell,
\begin{center}
{\it Discrete torsion is the choice of orbifold group action
on the $B$ field.}
\end{center}

More generally, in any theory containing fields with gauge
invariances, to define an orbifold it does not suffice to only
define the orbifold action on the base space.  One must also
specify the orbifold action on the fields.  The choice of orbifold
action on the fields is not unique, precisely because of the possibility
of combining the orbifold group action with a gauge transformation.

For vector fields, these degrees of freedom lead to orbifold Wilson lines.
For $B$ fields, these degrees of freedom lead to discrete torsion.

Put another way, discrete torsion can be understood in purely
mathematical terms, without any reference to string theory -- it
has nothing to do with string theory {\it per se}.
At the end of the day, one can calculate twisted sector phases,
orbifold group actions on D-branes, and so forth, so one can
certainly work out physical consequences of discrete torsion.
However, again, discrete torsion can ultimately be understood
in purely mathematical terms.

Phrased differently, discrete torsion can no longer be considered
to be mysterious -- a perfectly sensible, and purely mathematical,
description exists in \cite{dt1,dt2,dt3,cdt,hdt}.

In this article we shall give a brief overview of these
developments in understanding discrete torsion \cite{dt1,dt2,dt3,cdt,hdt}.

In section~\ref{orbWline} we discuss orbifold $U(1)$ Wilson lines,
which provide the prototype for our picture of discrete torsion.
In section~\ref{h2deriv} we briefly discuss how the group
$H^2(\Gamma, U(1))$ arises in describing orbifold group actions
on $B$ fields.  In section~\ref{twphase} we discuss how the twisted
sector phases of \cite{vafa1} can be derived.  In section~\ref{dougderiv}
we derive the projectivized orbifold group actions on D-brane worldvolumes
used by M.~Douglas in \cite{doug1,doug2} to describe discrete torsion
for D-branes.  In section~\ref{vwderiv} we outline how results
of C.~Vafa and E.~Witten \cite{vafaed} on the interplay between
discrete torsion and Calabi-Yau moduli can be understood in this 
framework.  In section~\ref{hdtrev} we briefly review results on
discrete torsion for perturbative heterotic strings.
Finally, in section~\ref{cfield} we briefly describe
analogues of this story for other tensor field potentials.

\section{Orbifold $U(1)$ Wilson lines}   \label{orbWline}

Our description of discrete torsion in \cite{dt1,dt2,dt3,cdt,hdt}
has many very close parallels with orbifold $U(1)$ Wilson lines.
For example, just on a naive level,
orbifold $U(1)$ Wilson lines are counted by
the group cohomology group\footnote{Where $\Gamma$ is the orbifold group.} 
$H^1(\Gamma, U(1))$, and discrete torsion
is associated with $H^2(\Gamma, U(1))$.
Also, orbifold $U(1)$ Wilson lines are technically far easier
and less subtle
to manipulate.  For these reasons, in \cite{dt1,dt2,dt3,cdt,hdt}
we repeatedly stress the parallels with orbifold $U(1)$
Wilson lines, and use orbifold $U(1)$ Wilson lines as
toy models for calculations involving $B$ fields.

How does one describe the action of an orbifold group
on a set of vector fields?
One way is to specify how the orbifold group acts on the total
space of the bundle; see for example \cite{dt1} for an example
of this approach.  Another (equivalent) way (described in \cite{dt3}) is
to describe pullbacks of the gauge fields and transition functions
in local coordinate patches.
For example, we discuss in \cite{dt3} how an orbifold group
action on a set of vector fields defined by $(A^{\alpha}, g_{\alpha \beta})$,
\begin{displaymath}
A^{\alpha} \: - \: A^{\beta} \: = \: d \log g_{\alpha \beta}
\end{displaymath}
can be written in the form
\begin{eqnarray*}
g^* A^{\alpha} & = & A^{\alpha} \: + \: d \log h^g_{\alpha} \\
g^* g_{\alpha \beta} & = & \left( g_{\alpha \beta} \right) \,
\left( h^g_{\alpha} \right) \,
\left( h^g_{\beta} \right)^{-1}
\end{eqnarray*}
where $g$ is an element of the orbifold group $\Gamma$,
for an appropriate collection of \v{C}ech cochains $\{ h^g_{\alpha} \}$.

How does the group $H^1(\Gamma, U(1))$ arise?
Not directly.  It can be shown that, in general, the set of
orbifold group actions is a {\it set}, and does not even possess
a group structure, so one cannot hope to understand orbifold
group actions directly from the calculation of some (cohomology) group.

However, it can be shown that the set of orbifold group actions,
although not itself a group, is acted upon by a group, namely $H^1(\Gamma,
U(1))$.
To be brief, given any one orbifold group action on a set of
vector fields (principal $U(1)$ bundle with connection),
one can get to any other action by combining that action with
a set of gauge transformations $\phi^g$.
These gauge transformations are required to respect the group
law, meaning that
\begin{displaymath}
\phi^{gh} \: = \: \phi^h \cdot h^* \phi^g
\end{displaymath}
and in order to preserve the gauge field, these gauge transformations
must be constant.  On a connected space, such a set of gauge transformations
defines a homomorphism $\Gamma \rightarrow U(1)$,
which is the same thing as an element of the group $H^1(\Gamma, U(1))$.
In other words, although the set of orbifold group actions does not
have a group structure, the difference between any two orbifold
group actions is described by an element of the group
$H^1(\Gamma, U(1))$.

Although the set of orbifold group actions does not in general
have a group structure, in special cases (in fact, the special
cases usually considered by physicists) one can impose
a group structure.  For example, if the bundle is topologically trivial,
then there is a notion of a canonical trivial action (which only acts
on the base, not the fiber).  In this special case,
one can describe any orbifold group action in terms of its
difference from the canonical trivial action, and so
(in this special case) one can speak of elements of $H^1(\Gamma, U(1))$
determining the orbifold group action.  In general, however, one must
be more careful.

Before going on to discuss $B$ fields, we should make a few
concluding remarks.  First, our analysis of orbifold group
actions on vector fields does not assume that $\Gamma$ acts
freely.  Our analysis also does not assume that $\Gamma$ is
abelian.  Finally, our analysis also does not assume
that the bundle on which the gauge fields live is topologically
trivial, although if it is not trivial, one must be careful
to check that any orbifold group actions exist, which
our analysis assumes.

\section{$H^2(\Gamma, U(1))$}    \label{h2deriv}

In this section we shall give a pedagogical outline of
orbifold group actions on $B$ fields.  Our discussion here is not
meant to be rigorous; however, rigorous versions do exist.
In particular, reference \cite{dt2} contains a completely rigorous description
of orbifold group actions on $B$ fields, and reference \cite{dt3}
contains a substantially simplified version.

In general terms, the same analysis holds for $B$ fields
as for vector fields.  One can describe elements of the set of
orbifold group actions on the $B$ fields, and
just as the difference between any two orbifold group actions
on a $U(1)$ gauge field is a set of gauge transformations $\phi^g$,
the difference between any two orbifold group transformations
on a $B$ field is a set of gauge transformations.

What is a gauge transformation of a $B$ field?
Locally, recall a gauge transformation of a $B$ field looks like
\begin{displaymath}
B \: \mapsto \: B \: + \: d \Lambda
\end{displaymath}
for $\Lambda$ some one-form.  We shall omit the details here
(see instead \cite{dt1,dt2}), however the reader may correctly
guess that globally, a gauge-transformation of a $B$ field
is determined by a principal $U(1)$ bundle with connection
-- i.e., a set of $U(1)$ gauge fields $\Lambda$.

In other words, to specify the difference between two orbifold
group actions on a $B$ field, we specify a set of principal $U(1)$
bundles $T^g$ with connection $\Lambda(g)$.
Also, just as we had to restrict to constant gauge transformations
of $U(1)$ gauge fields in order to preserve the connection and
count orbifold $U(1)$ Wilson lines, here we must demand that
the connection $\Lambda(g)$ on $T^g$ be flat.

Now, there are two important subtleties in this description.

First, if $\Lambda$ and $\Lambda'$ are two connections that
differ by a (bundle) gauge transformation, then they define
the same gauge transformation of the $B$ field.
After all, the $B$ field transforms as $B \mapsto B + d \Lambda$,
so we can add any exact form to $\Lambda$ without changing
the gauge transformation of $B$.
Globally, if $(T^g, \Lambda(g))$ and $(T'^g, \Lambda'(g))$
are two principal $U(1)$ bundles with connection
that differ by a connection-preserving bundle isomorphism,
then they define the same gauge transformation of the $B$ field.

Second, not any set $(T^g, \Lambda(g))$ will do -- they must
be well-behaved with respect to group composition.
In other words, we must demand an analogue of the constraint
\begin{equation}    \label{phiconstr}
\phi^{g_1 g_2} = \phi^{g_2} \, \cdot \, g_2^* \phi^{g_1} 
\end{equation}
that we imposed on the gauge-transformations relating orbifold group
actions on $U(1)$ gauge fields.
Instead of $U(1)$ valued functions $\phi^g$, we now have
bundles $T^g$.  The correct analogue of equation~(\ref{phiconstr})
is that the bundle $T^h \otimes h^* T^g$ must be isomorphic
to the bundle $T^{gh}$, and the same isomorphism must preserve
the connection.

Let $\omega^{g,h}: T^h \otimes h^* T^g \rightarrow T^{gh}$
denote that connection-preserving isomorphism.
Now, by composing the isomorphisms $\omega$, we can construct
two different isomorphisms between
$T^{g_3} \otimes g_3^* \left( T^{g_2} \otimes g_2^* T^{g_1} \right)$
and $T^{g_1 g_2 g_3}$; to be consistent, these two isomorphisms must
be the same.

To summarize, we have stated that any two
orbifold group actions on a set of $B$ fields differ by a collection
of principal $U(1)$ bundles $T^g$ with flat connection $\Lambda(g)$,
together with connection-preserving bundle morphisms
$\omega^{g_1, g_2}: T^{g_2} \otimes g_2^* T^{g_1} \rightarrow
T^{g_1 g_2}$, such that the following diagram commutes:
\begin{equation}    \label{omegacocycle}
\begin{array}{ccc}
T^{g_3} \otimes g_3^* \left( \, T^{g_2} \otimes g_2^* T^{g_1} \right)
& \stackrel{ \omega^{g_1, g_2} }{ \longrightarrow } &
T^{g_3} \otimes g_3^* T^{g_1 g_2} \\
\makebox[0pt][r]{ $\scriptstyle{  \omega^{g_2, g_3} }$} \downarrow
& & \downarrow \makebox[0pt][l]{
$\scriptstyle{ \omega^{g_1 g_2, g_3} }$ } \\
T^{g_2 g_3} \otimes (g_2 g_3)^* T^{g_1} &
\stackrel{ \omega^{g_1, g_2 g_3} }{\longrightarrow } &
T^{g_1 g_2 g_3}
\end{array}
\end{equation}
Also, we are free to replace any of the bundles $T^g$ with
isomorphic bundles.  Let $T'^g$ be another set of principal $U(1)$ bundles
with connection $\Lambda'(g)$, and $\kappa_g: T^g \rightarrow T'^g$
connection-preserving bundle isomorphisms,
then we can replace the data given above with the collection
$(T'^g, \Lambda'(g), \omega'^{g_1, g_2})$, where the $\omega'$ are
given by
\begin{equation}    \label{omegacobound}
\omega'^{g_1, g_2} \: \equiv \:
\kappa_{g_1 g_2} \circ \omega^{g_1, g_2} \circ
\left( \kappa_{g_2} \otimes g_2^* \kappa_{g_1} \right)^{-1}
\end{equation}

How does $H^2(\Gamma, U(1))$ arise in this description?
Consider the special case that all the bundles $T^g$ are topologically
trivial, and the connections $\Lambda(g)$ are gauge-trivial.
Without loss of generality, we can map each $T^g$ to the canonical
trivial bundle (whose transition functions are identically 1)
and gauge-transform the connections $\Lambda(g)$ to be identically
zero, not just gauge-trivial.  Then the maps $\omega^{g,h}$ 
can be interpreted as gauge transformations (of the canonical
trivial bundle), and furthermore constant gauge transformations
(since they must preserve the connection, which is the same -- zero --
on both sides).  Commutivity of diagram~(\ref{omegacocycle})
then becomes the group 2-cocycle condition.
So, in this choice of gauge, the maps $\omega^{g,h}$ are explicitly
representatives of an element of $H^2(\Gamma, U(1))$.

In addition, there are residual gauge transformations -- we can 
gauge-transform each of the bundles $T^g$ by a constant gauge transformation,
without making the connection $\Lambda(g)$ nonzero.  From
equation~(\ref{omegacobound}) we see that such gauge transformations
of the bundles $T^g$ change the $\omega^{g,h}$ by a group coboundary.
So, $B$ field gauge transformations of this form
are classified by elements of $H^2(\Gamma, U(1))$.

In addition to these $B$ field gauge transformations,
which yield elements of $H^2(\Gamma, U(1))$, there can be
additional $B$ field gauge transformations, a fact which does not
seem to have been noticed previously in the physics literature.
These are discussed in more detail in \cite{dt3}.

\section{Twisted sector phases}   \label{twphase}

How do the twisted sector phases of \cite{vafa1} arise in this
picture?
In a nutshell, they arise because of the sigma model term
$\int B$, and are precise analogues of Wilson loops.

\begin{figure}
\centerline{\psfig{file=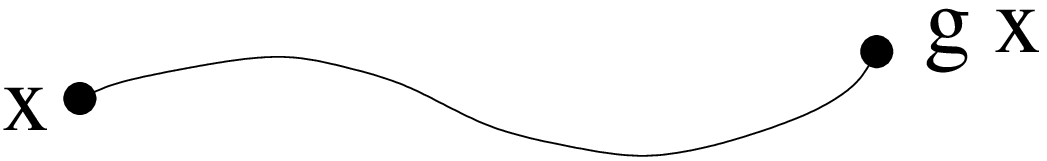,width=2in}}
\caption{  \label{figWline}
Lift of closed loop to covering space.}
\end{figure}

Consider Wilson loop on a quotient space for a moment.
Suppose on the covering space, the Wilson loop becomes an open line,
with ends identified by the orbifold group action, as illustrated
in figure~(\ref{figWline}).  The Wilson loop on the quotient space
is not merely
\begin{displaymath}
\exp \, \left( \, \int^{gx}_x A \, \right)
\end{displaymath}
but also picks up a factor from a gauge transformation relating
the gauge field $A$ at $x$ and $gx$.  In other words, the Wilson loop
on the quotient space is given by
\begin{displaymath}
\varphi \, \exp \, \left( \, \int^{gx}_x A \, \right)
\end{displaymath}
for some $\varphi$.

In \cite{dt1,dt3} we point out that the same considerations for
Wilson surfaces $\exp \left( \int B \right)$ give rise to 
the twisted sector phases of \cite{vafa1}.

\begin{figure}   
\centerline{\psfig{file=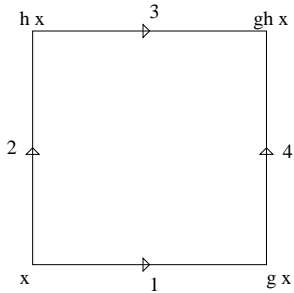,width=1.5in}}
\caption{  \label{fig1loop}
A twisted sector
contribution to the one-loop partition function.}
\end{figure}

For simplicity\footnote{So we can describe any orbifold group
action in terms of a set of global gauge transformations distinguishing
it from a canonical trivial action.}, assume $B$ is topologically trivial.
Consider a polygon on the covering space as shown in figure~(\ref{fig1loop}),
describing a 
contribution to a one-loop twisted sector partition function.
Naively, one might write the Wilson surface on the quotient space
as merely $\exp \left( \int B \right)$, but just as for Wilson loops,
this is wrong because it fails to take into account gauge transformations
along the boundary.

As discussed earlier, a nontrivial orbifold group action on
a (topologically trivial) $B$ field is determined by a set of
$B$ field gauge transformations -- that is, a set of principal
$U(1)$ bundles $T^g$ with connection $\Lambda(g)$, and connecting
maps $\omega^{g,h}$.

To take into account gauge transformations of the $B$ field
along the boundary, one naively would correct the factor
$\exp \left( \int B \right)$ by adding a factor
\begin{displaymath}
\exp \, \left( \, \int^{hx}_x \Lambda(g) \: - \:
\int^{gx}_x \Lambda(h) \, \right)
\end{displaymath}

As discussed in \cite{dt3}, this is still not quite right, if for no
other reason than it is not invariant under gauge transformations of
the bundles $T^g$.  To get a gauge-invariant result we must take into
account the corners, which leads one to
\begin{displaymath}
\left( \omega^{g,h}_x \right) \,
\left( \omega^{h,g}_x \right)^{-1} \,
\exp \, \left(  \, \int^{hx}_x \Lambda(g) \: - \:
\int^{gx}_x \Lambda(h) \, \right)
\end{displaymath}

Since this expression is gauge-invariant, we can evaluate it in any
convenient gauge.  For $B$ field gauge transformations corresponding
to elements of $H^2(\Gamma, U(1))$,
a particularly convenient gauge is one in which
all $\Lambda(g) \equiv 0$ and the $\omega^{g,h}$ are constant.
Working in this gauge, it is now clear that for such
$B$ field gauge transformations, the expression above is
equal to the constant
\begin{displaymath}
\left( \omega^{g,h} \right) \,
\left( \omega^{h,g} \right)^{-1}
\end{displaymath}
which is precisely the phase factor given in \cite{vafa1}.
Since this phase factor is constant, all contributions to the twisted
sector partition function $Z_{(g,h)}$ are multiplied by the same factor,
and so the effect is to multiply $Z_{(g,h)}$ by this phase.

In \cite{dt3} we also derive the phase factors appearing in
higher-loop string orbifold partition functions.

\section{M. Douglas's orbifold actions and D-branes}   \label{dougderiv}

In \cite{doug1,doug2}, it was claimed that in terms of orbifold
group actions on D-branes, discrete torsion is realized in terms of
a projectivized action of the orbifold group.  

In \cite{dt3} we point out that this orbifold group action on D-brane
worldvolumes is a natural consequence of the fact that the
``bundle'' on a D-brane is actually twisted by the $B$ field.
Specifically, since gauge transformations $B \mapsto B + d \Lambda$
induce $A \mapsto A + \Lambda$, the $B$ field and the D-brane
gauge fields are correlated.  For a topologically trivial $B$
field, for example, the transition functions for the ``bundle''
on the D-brane do not close as
\begin{displaymath}
g_{\alpha \beta} \, g_{\beta \gamma} \, g_{\gamma \alpha} \: = \: I
\end{displaymath}
but rather close only up to a \v{C}ech cocycle $h_{\alpha \beta \gamma}$
defining the
characteristic class of the $B$ field:
\begin{displaymath}
g_{\alpha \beta} \, g_{\beta \gamma} \, g_{\gamma \alpha} \: = \:
h_{\alpha \beta \gamma} I
\end{displaymath}

Once one understands orbifold group actions on $B$ fields,
it is then straightforward to work out orbifold group
actions on these ``twisted bundles'' appearing on D-brane worldvolumes,
and one rapidly recovers the projectivized orbifold group
action described in \cite{doug1,doug2}.

\section{Vafa-Witten}   \label{vwderiv}

Reference \cite{vafaed} studied how discrete torsion
affects Calabi-Yau moduli.  Specifically, the authors of that
paper studied deformations of toroidal orbifolds in which
discrete torsion had been turned on.  In general terms,
they found that all nontrivial K\"ahler moduli were lifted,
as well as many of the complex structure moduli.

In \cite{dt3} we point out that these behaviors have 
a natural understanding in our picture.  
Briefly, turning on discrete torsion
means that the quotient space has nontrivial Wilson surfaces
$\exp \left( \int B \right)$.  Just as turning on orbifold
Wilson lines on quotients ${\bf C}^2/\Gamma$, for example,
forces resolutions to have nonzero curvature supported on
exceptional divisors, so nonzero Wilson surfaces forces one
to need nonzero curvature $H$ on any full resolution.

A blowup or small resolution of a toroidal orbifold
tends to only introduce even-dimensional cycles,
leaving no natural place for the $B$ field curvature $H$ to live.
Therefore, one is naturally led to believe that nontrivial
K\"ahler deformations are obstructed by the demands of
consistency, and indeed this is what was observed in 
\cite{vafaed}.

Complex structure resolutions, on the other hand, often introduce
3-cycles.  So one could deform the complex structure of the toroidal
orbifold, and naturally be led to a space with nonzero curvature $H$.
However, nonzero $H$ on a (K\"ahler) Calabi-Yau breaks supersymmetry,
so any complex structure deformation that completely resolves the
singularity would also be obstructed, as was noted in \cite{vafaed}.

What deformations remain?  One can still consistently deform
the complex structure, but one cannot expect to completely
resolve the space (at least, not to a Calabi-Yau).
This is precisely what was found in \cite{vafaed}.

A somewhat more detailed examination of how \cite{vafaed}
can be naturally understood in the framework we have presented
can be found in \cite{dt3}.

\section{Perturbative heterotic strings}    \label{hdtrev}

So far we have reviewed orbifold group actions on $B$ fields
whose curvature $H$ is closed, in the sense that $d H = 0$.
Unfortunately, the $B$ fields of string
theory do not always have this property.  The most prominent exception
is the $B$ field of perturbative heterotic string theory.
In this case, the curvature $H$ is a (globally-defined) three-form
such that $d H = \mbox{Tr } F \wedge F - \mbox{Tr } R \wedge R$,
and so (except for the special case of
the old-fashioned ``standard embedding''), $d H \neq 0$ for the
heterotic $B$ field.

We examined orbifold group actions on the heterotic $B$ field
in detail in \cite{hdt}.  To be brief, we found that, for fixed
orbifold group action on the gauge and tangent bundle,
the difference between any two orbifold group actions on
a heterotic $B$ field was defined by the same data as for a type II
$B$ field.  In other words, the difference between any two orbifold
group actions on a heterotic $B$ field is defined by data
$(T^g, \omega^{g_1, g_2})$, where the $T^g$ are 
principal $U(1)$-bundles with connection (one for each $g \in \Gamma$),
and the $\omega^{g_1, g_2}$ are connection-preserving bundle isomorphisms,
subject to the same equivalence relations as before.

Thus, for heterotic $B$ fields as well as type II $B$ fields,
we find the same occurrence of $H^2(\Gamma, U(1))$, as well as
twisted-sector phases.

\section{Analogues for $C$ fields}  \label{cfield}

We have repeatedly stressed that orbifold group actions on any
field with gauge invariances are not unique, because one can
combine the orbifold group action with a gauge transformation.
For vector fields, this ambiguity is known as ``orbifold Wilson lines,''
and we have explained in \cite{dt1,dt2,dt3} how for $B$ fields
this ambiguity is discrete torsion.

String theory has other fields beyond vector fields and $B$ fields;
what degrees of freedom does one find in defining orbifold group
actions on these other fields?

In reference \cite{cdt} we explore this matter for the M-theory
three-form.  In order to get concrete results, we make the
slightly simplistic assumption that the $C$ field can be
understood as a connection on a 2-gerbe.
When we repeat the calculations of \cite{dt3}, we find 
that part of the difference between orbifold group actions
is classified by $H^3(\Gamma, U(1))$, though there can be additional
orbifold group actions also.

One might ask what the analogue of the twisted sector phases
of \cite{vafa1} are for membranes.  The answer is as follows
\cite{dt3}:  given a ``twisted sector'' on $T^3$ defined by a commuting
triple $(g_1, g_2, g_3)$, under an orbifold group action
on a trivial $C$ field determined by
an element of $H^3(\Gamma, U(1))$, the factor 
\begin{displaymath}
\exp \, \left( \, \int C \, \right)
\end{displaymath}
picks up a factor of
\begin{displaymath}
\left( \omega^{g_1, g_2, g_3} \right) \,
\left( \omega^{g_2, g_1, g_3} \right)^{-1} \,
\left( \omega^{g_3, g_2, g_1} \right)^{-1} \,
\left( \omega^{g_3, g_1, g_2} \right) \,
\left( \omega^{g_2, g_3, g_1} \right) \,
\left( \omega^{g_1, g_3, g_2} \right)^{-1}
\end{displaymath}
from gauge transformations at the boundaries,
where $\omega^{g_1, g_2, g_3}$ is a group 3-cocycle.
This expression is invariant under changing the cocycle by
a coboundary (as it must be, to associate a well-defined phase
to an element of $H^3(\Gamma, U(1))$), and is also invariant
under the natural $SL(3,{\bf Z})$ action on $T^3$.

Admittedly the analysis in \cite{cdt} is slightly naive -- it
is really just an analysis of orbifold group actions on 2-gerbes,
and does not take much physics into account.
For example, it ignores the flux-quantization-shifting effect
described in \cite{fluxquant}, which almost certainly will
have an effect here.

However, although the analysis is naive, it does highlight
the fact that one does expect an analogue of discrete torsion
for the M-theory $C$ field, and gives a first rough idea
of what those degrees of freedom should look like.

\section{Conclusions}

In this short article we have given a brief general overview of
the basic understanding of discrete torsion described in
\cite{dt1,dt2,dt3,cdt}.
Those references explain that discrete torsion is simply the 
choice of orbifold group action on the $B$ field; all the basic
properties of discrete torsion, from its appearance as twisted sector
phases \cite{vafa1} to M. Douglas's description of discrete torsion
for D-branes \cite{doug1,doug2} follows from this description.

\section{Acknowledgements}

We would like to thank  P.~Aspinwall, A.~Knutson, D.~Morrison,
and R.~Plesser for useful conversations.

This research was partially supported by National Science Foundation
grant number DMS-0074072.

\end{document}